\documentstyle[twoside,fleqn,espcrc2]{article}

\newcommand{\AmS}{{\protect\the\textfont2
  A\kern-.1667em\lower.5ex\hbox{M}\kern-.125emS}}

\input epsf
\title{Critical exponents in abelian projected $SU(2)$ QCD} 
\author{Shun-ichi Kitahara\address{%
        Jumonji University, Niiza, Saitama 352, Japan
        },%
        Shinji Ejiri\address{%
        Department of Physics, Kanazawa University, 
        Kanazawa 920-11, Japan
        },%
        Tsuneo Suzuki\addtocounter{address}{-1}\addressmark ~
        and 
        Koji Yasuta\addtocounter{address}{-1}\addressmark
}

\begin{document}

\begin{abstract}
The critical properties of the abelian Polyakov loop 
and the Polyakov loop in terms of Dirac string are studied 
in finite temperature abelian projected $SU(2)$ QCD.
The critical point and the critical exponents are determined from 
each Polyakov loop in the maximally abelian gauge 
using the finite-size scaling analyses.
Those critical points and exponents are in good agreement with 
those from non-abelian Polyakov loops.
\end{abstract}
\newpage

\maketitle

\section{Introduction} 

Abelian projected QCD is regarded as an abelian theory 
with electric charges and monopoles.
Dual Meissner effect based on 
the condensation of the monopoles can be considered 
as the mechanism of confinement of quarks.
This picture is likely to be  realized at least 
in the maximally abelian gauge: 
the value of the string tension and the behavior of Polyakov loops 
are reproduced by the abelian link fields and by the monopoles 
(Dirac strings) 
\cite{hio,shiba4,suzu94a};
the effective monopole action is calculated 
and it indicates that QCD is always in the monopole condensed phase
from the comparison of the energy and the entropy 
\cite{shiba2a}.

\begin{figure}
\epsfxsize=70mm\epsfysize=50mm
 \vspace{-20pt}
 \begin{center}
 \leavevmode
\epsfbox{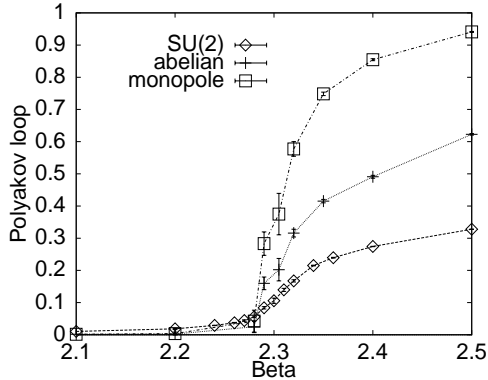}
 \end{center}
 \vspace{-37pt}
\caption{
Non-abelian, abelian and 
monopole Polyakov loops versus $\beta$ on $16^3\times 4$ 
$SU(2)$ lattice.
\label{POLYAKOVLOOP}
}
\vspace{-15pt}
\end{figure}

Figure \ref{POLYAKOVLOOP}\cite{suzu94a} shows 
the abelian and the monopole Polyakov loops defined later appear 
to be good order parameters.
However, those curves seem to have different slopes.
Their absolute values in the deconfinement phase are also different.
The critical behavior of 4-dimensional $SU(2)$ lattice gauge theory 
is shown  to be the same as that of 3-dimensional $Z_2$ theory.
Since the abelian and the monopole Polyakov loops seem 
to be good order parameters, it is interesting to evaluate 
critical exponents and critical points exactly from these 
Polyakov loops.
What kind of universality class will appear from these abelian quantities?

\section{Definitions of Polyakov loops } 

After abelian projection is over, we can define 
abelian Polyakov loops\cite{hio} 
written in terms of abelian link variables $u_\mu(x,t)$.
We separate $u_\mu(x,t)$ from 
gauge-fixed link variables $\widetilde{U}_\mu (x,t)$:
$\widetilde{U}_\mu (x,t)=A_\mu(x,t)u_\mu(x,t)$.
We  define an abelian Polyakov loop 
\begin{eqnarray}
 P_{abel}(x_0)=\mbox{Re\,}[\exp (i\sum\theta_4(x,t)J_4(x,t))] 
 \label{eqn:ABELPLOOP} .
\end{eqnarray}
Here $J_\mu(x,t)=\delta_{\mu,4}\delta_{x,x_0}$ and $\theta_\mu (x,t)$ are the 
angle variables of $u_\mu(x,t)=\exp(i\theta_\mu (x,t)\sigma_3)$, 
where $\sigma_3$ is a Pauli matrix.

The abelian Polyakov loop can be decomposed into two parts: 
a monopole part and a photon part\cite{suzu94a}.
The abelian field strength 
$  f_{\mu\nu} = \partial_{\mu}\theta_{\nu} - %
    \partial_{\nu}\theta_{\mu} $
can be separated into two parts:
$f_{\mu\nu}=\bar{f}_{\mu\nu}+2\pi n_{\mu\nu}$,
where $n_{\mu\nu}$ is an integer and $\bar{f}_{\mu\nu}\in [-\pi,\pi)$.
Then, rewriting Eq.(\ref{eqn:ABELPLOOP}), we get
\vspace{.3cm}
\begin{eqnarray*}
  P_{abel}(x_0)&=&\mbox{Re\,}[P_1(x_0)\cdot P_2(x_0)],\\
 P_1(x_0) &=&
    \exp (-i\sum D(x-x',t-t') \\ & & \times
    \partial'_{\nu}\bar{f}_{\nu 4}(x',t')J_4(x,t)) ,\\
 P_2(x_0) &=&
    \exp (-2\pi i\sum D(x-x',t-t') \\ & & \times
    \partial'_{\nu}n_{\nu 4}(x',t')J_4(x,t)) .
\end{eqnarray*}
Here 
$D(x,t)$ is a lattice Coulomb propagator which satisfies 
$\partial'_\mu\partial_\mu D(x,t)=-\delta_{x,0}\delta_{t,0} $.
The monopole Polyakov loop, $P_{mono}(x_0)=\mbox{Re\,}P_1(x_0)$ is 
composed of Dirac strings of monopoles.
Suzuki et al.\cite{suzu94a} have indicated that 
$P_{abel}(x_0)$ and $P_{mono}(x_0)$ vanish in the confinement phase, 
whereas
$P_{photon}=\mbox{Re\,}P_2(x_0)$ is finite 
at the range from $\beta=$2.1 to 2.5 
and did not change drastically around the critical point.

\section {Finite-size scaling theory} 

We calculated the critical exponent of the non-abelian, 
the abelian and the monopole Polyakov loops 
from a finite-size scaling theory.
The singular part of the free energy density on 
$N_s^3\times N_t$ lattice has the following form:
\begin{eqnarray*}
 f_s(x,h,N_s)=N_s^{-d}Q_s(xN_s^{1/\nu},hN_s^{(\beta+\gamma)/\nu}
,g_\omega N_s^{-\omega}),
\end{eqnarray*}
where $x=(T-T_c)/T_c$. Here the action contains the term $hN_s^{d}L$ 
($L$ denotes the magnetization) 
and only the largest irrelevant exponent $(-\omega)$ is taken 
into account.
By differentiating $f_s$ with respect to $h$ at $h=0$, we get 
\begin{eqnarray*}
 L(x,N_s)&=&N_s^{-\beta/\nu}Q_L (xN_s^{1/\nu},g_\omega N_s^{-\omega}) ,\\
 \chi(x,N_s)&=&N_s^{\gamma/\nu}Q_\chi (xN_s^{1/\nu},g_\omega N_s^{-\omega}) ,\\
 g_r(x,N_s)&=&Q_{g_r} (xN_s^{1/\nu},g_\omega N_s^{-\omega}) ,
\end{eqnarray*}  
where $\langle L \rangle$, $\chi$ and $g_r$ are 
order parameter, susceptibility and 4-th cumulant, respectively.
Expanding those equations with respect to $x$, we have
\begin{eqnarray}
  L(x=0,N_s)&=&N_s^{-\beta/\nu}(c_0+c_3N_s^{-\omega}) %
  \label{eqn:Lomega} ,\\
  \chi(x=0,N_s)&=&N_s^{\gamma/\nu}(c_0+c_3N_s^{-\omega}) %
  \label{eqn:chiomega} ,\\
  g_r(x=0,N_s)&=& g_r^\infty+c_3N_s^{-\omega} , \label{eqn:grfit}
\end{eqnarray}
at $x=0$.
The critical point can be defined as that point where a fit to the leading 
$N_s$-behavior has the least minimal $\chi^2$\cite{engels}.
Actually, leading $N_s$-behavior of Eq.(\ref{eqn:Lomega}) 
and of Eq.(\ref{eqn:chiomega}) is given by
\begin{eqnarray}
  \ln L(x=0,N_s)&=&-({\beta}/{\nu})\ln N_s + \ln c_0 \label{eqn:Lfit} ,\\
  \ln \chi(x=0,N_s)&=&({\gamma}/{\nu})\ln N_s + \ln c_0 \label{eqn:chifit} ,
\end{eqnarray}
as in ref.\cite{engels}.
From the fits to 
Eqs.(\ref{eqn:grfit}), (\ref{eqn:Lfit}) and (\ref{eqn:chifit}), 
we can find the position of critical point $\beta_c$,
and then obtain the values of $\beta/\nu$,$\gamma/\nu$ and $g_r^{\infty}$ 
at $\beta_c$ simultaneously.
We also considered the derivatives of the observables 
with respect to $x$. 
The leading $N_s$-behavior of each derivatives at the critical point is 
obtained similarly:
\begin{eqnarray}
  \ln\frac{\partial O}{\partial x}(x=0,N_s)
  = \rho \ln N_s + \ln c_0 \label{eqn:dLfit} .
\end{eqnarray}
Here $O$ is $L$, $\chi$ and $g_r$ with 
($\rho=(1-\beta)/\nu$, $(1+\gamma)/\nu$ and $1/\nu$). 

\begin{figure}
\epsfxsize=60mm\epsfysize=50mm
 \vspace{-5pt}
 \begin{center}
 \leavevmode
\epsfbox{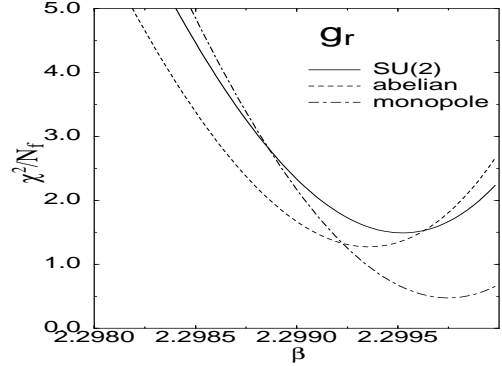}
 \end{center}
 \vspace{-37pt}
\caption{
$\chi^2/N_f$ from $g_r$ fits 
versus $\beta$ in the non-abelian, the abelian and the monopole cases.
The number of degrees of freedom, $N_f$ is 2.
\label{CHISQ}
}
\vspace{-18pt}
\end{figure}
\begin{table*}[hbt]
\begin{center}
\begin{tabular}{|l|l|l|l||l|l||l|}
\hline
 &{$SU(2)$}&{abel}&{mono}& {Engels et al.\cite{engels}} &
 {Ising\cite{ferr2}} & $U(1)$\cite{svet} \\  
\hline
$\beta/\nu$    & 0.504(18)& 0.485(22)& 0.528(64)& 0.525(8) & 0.518(7)& \\
$(1-\beta)/\nu$& 1.117(27)& 1.138(10)& 1.091(84)& 1.085(14)& 1.072(7)& \\
$\nu  $   & 0.617(16)& 0.616(12)& 0.617(57)& 0.621(6) & 0.6289(8)& 0.67 \\
$\beta$   & 0.311(19)& 0.299(19)& 0.326(69)& 0.326(8) & 0.3258(44)& 0.35 \\
\hline
$\gamma/\nu$   & 1.977(29)& 2.025(34)& 1.991(88)& 1.944(13)& 1.970(11) & \\
$(1+\gamma)/\nu$&3.600(38)& 3.646(44)& 3.608(93)& 3.555(15)& 3.560(11) & \\
$\nu  $   & 0.616(25)& 0.617(29)& 0.618(68)& 0.621(8) & 0.6289(8) & \\
$\gamma$  & 1.218(68)& 1.249(81)& 1.23(19) & 1.207(24)& 1.239(7)  & 1.32 \\
\hline
$\gamma/\nu+2\beta/\nu$ %
                  & 2.985(47)& 2.995(56)& 3.05(15)& 2.994(21)& 3.006(18)& \\
\hline
$-g_r^\infty$     & 1.447(41)& 1.438(42)& 1.438(41)& 1.403(16)& 1.41 & \\
$\nu$     & 0.633(13)& 0.621(14)& 0.600(13)& 0.630(11)& 0.6289(8) & \\
\hline
\end{tabular}
\end{center}
\caption{%
The critical exponents calculated from various Polyakov loops.
 \label{table:INDEX}
}
\end{table*}

\section{Results and Discussions} 

We performed the numerical calculations on 
$N_s^3\times 4$ lattices, where $N_s=$8,12,16 and 24.
The standard $SU(2)$ Wilson action was adopted and abelian link valuables 
were defined in maximally abelian gauge.
We calculated the following observables:
\begin{eqnarray*}
  L&=&\sum_xP(x)/{N_s^3}  ,\\
  \chi&=&N_s^3(\langle L^2\rangle - \langle L\rangle^2) , \\
  g_r&=&{\langle L^4\rangle}/{{\langle L^2\rangle}^2}-3 ,
\end{eqnarray*}%
where 
$P(x)$ denotes $P_{SU(2)}(x)$, $P_{abel}(x)$ and $P_{mono}(x)$. 
The values of observables at various $\beta$ are needed 
in order to calculate the derivatives 
with respect to $x$, where $x=(\beta-\beta_c)/\beta_c$.
Then, we used the density of state method(DSM)\cite{ferr}.
First we performed Monte-Carlo simulations at $\beta_0$=2.2988, 
and then the expectation values of the observables 
in the vicinity of $\beta_0$ were obtained using DSM. 
$L$, $\chi$ and $g_r$ at $\beta_0$ were calculated every 50 sweeps 
after 2000 thermalization sweeps.
The number of samples was 100000,
except on $24^3\times 4$ lattice (47000 in the case).
The errors were determined according to the 
Jackknife method dividing the entire sample into 10 blocks 
(4 blocks on $24^3\times 4$ lattice).

We estimated the critical point $\beta_c$ from $\chi^2$ method\cite{engels}.
The data of our DSM results are fitted to 
Eq.(\ref{eqn:grfit})-(\ref{eqn:chifit}) and 
Eq.(\ref{eqn:dLfit})
at each $\beta$.
The number of input data is 2 and that of fit parameters is 2 
($\omega$ in Eq.(\ref{eqn:grfit}) is fixed to 1 in accordance 
with Engels et al.\cite{engels}).
Figure \ref{CHISQ} describes the typical curves of 
$\chi^2/N_f$ versus $\beta$.

Averaging 
the obtained minimal positions of $\chi^2/N_f$, we get
\begin{eqnarray*}
   \beta_c^{SU(2)}      &=& 2.29940(20) , \\
   \beta_c^{abel} \;\;\,&=& 2.99962(26) , \\
   \beta_c^{mono} \;    &=& 2.29971(23) . 
\end{eqnarray*}
Those critical points are very close to each other.

Table \ref{table:INDEX} lists the critical exponents on each critical point 
in the non-abelian, the abelian and the monopole case.
We get the following results:
\begin{enumerate}
\item
The critical exponents in the abelian and the monopole case 
are in agreement with non-abelian exponents within the statistical 
error.
\item
Those critical exponents agree with those of $Z_2$ rather than 
those of $U(1)$. 
\item
Hyperscaling relations are well satisfied.
\item
Non-abelian exponents obtained are consistent with those 
of Engels et al.\cite{engels}.

\end{enumerate}

The first and the second results indicate the abelian (monopole) 
dominance in quark confinement.


\begin{thebibliography}{99}
\bibitem{hio} S. Hioki $et$ $al.$, Phys. Lett. {\bf B272} (1991) 326.
\bibitem{shiba4} H. Shiba and T. Suzuki, 
Phys. Lett. {\bf B333} (1994) 461. 
\bibitem{suzu94a} T. Suzuki $et$ $al.$,  
Phys. Lett. {\bf B347} (1995) 375. 
\bibitem{shiba2a} H. Shiba and T. Suzuki, 
Phys. Lett. {\bf B351} (1995) 519. 
\bibitem{engels} J. Engels $et$ $al.$, 
University of Bielefeld preprint, BI-TP 95/29, 1995.
\bibitem{ferr} A.M. Ferrenberg and R.H. Swendsen, 
Phys. Rev. Lett. {\bf 63} (1989) 1195.
\bibitem{ferr2} A.M. Ferrenberg and D.P. Landau, 
Phys. Rev. {\bf B44} (1991) 5081.
\bibitem{svet} B. Svetitsky 
Phys. Pep. {\bf 132} (1986) 1.
\end{thebibliography}
\end{document}